# Ultrafast electron calorimetry uncovers a new long-lived metastable state in 1$T$-TaSe$_2$ mediated by mode-selective electron-phonon coupling

Xun Shi[1]*†, Wenjing You[1]†, Yingchao Zhang[1]†, Zhensheng Tao[1]‡, Peter M. Oppeneer[2], Xianxin Wu[3], Ronny Thomale[3], Kai Rossnagel[4,5,6], Michael Bauer[4], Henry Kapteyn[1], Margaret Murnane[1]

[1]*Department of Physics and JILA, University of Colorado and NIST, Boulder, Colorado 80309, USA*

[2]*Department of Physics and Astronomy, Uppsala University, Box 516, 75120 Uppsala, Sweden*

[3]*Institut fur Theoretische Physik und Astrophysik, Julius-Maximilians-Universitat Wurzburg, 97074 Wurzburg, Germany*

[4]*Institute of Experimental and Applied Physics, Kiel University, D-24098 Kiel, Germany*

[5]*Deutsches Elektronen-Synchrotron DESY, D-22607 Hamburg, Germany*

[6]*Ruprecht Haensel Laboratory, Kiel University and DESY, D-24098 Kiel and D-22607 Hamburg, Germany*

*Corresponding author. Email: Xun.Shi@colorado.edu

†These authors contributed equally to this work.

Quantum materials represent one of the most promising frontiers in the quest for faster, lightweight, energy efficient technologies. However, their inherent complexity and rich phase landscape make them challenging to understand or manipulate in useful ways. Here we present a new ultrafast electron calorimetry technique that can systematically uncover new phases of quantum matter. Using time- and angle-resolved photoemission spectroscopy, we measure the dynamic electron temperature, band structure and heat capacity. We then show that this is a very sensitive probe of phase changes in materials, because electrons react very quickly, and moreover generally are the smallest component of the total heat capacity. This allows us to uncover a new long-lived metastable state in the charge density wave material 1$T$-TaSe$_2$, that is distinct from all of the known equilibrium phases: it is characterized by a significantly reduced effective heat capacity that is only 30% of the normal value, due to selective electron-phonon coupling to a subset of phonon modes. As a result, significantly less energy is required to melt the charge order and transform the state of the material than under thermal equilibrium conditions.

*‡Current address: State Key Laboratory of Surface Physics, Department of Physics, Fudan University, Shanghai*

**Introduction**

The discovery of new phases of matter has a long history of providing enhanced functionality, for example for phase change memories. Moreover, the critical need for faster, smaller and more energy-efficient nanotechnologies means that metastable phases of matter are a new frontier for functional materials(*1–10*). Correlated materials exhibit rich phase diagrams due to the coupling of charge, lattice, spin and orbital degrees of freedom. These intertwined interactions can give rise to intriguing phenomena such as superconductivity, magnetism and density waves(*11, 12*). Specifically, charge density wave (CDW) states, in which both the electron density and atomic lattice are modulated periodically with a characteristic wave vector ($q_{CDW}$), are believed to arise from electron-phonon coupling, sometimes combined with, or even mediated by electron-electron interactions(*11*). Due to the complexity of these interactions, the nature and origin of different phases and their transitions in strongly correlated materials are still under debate. In addition, the potential presence of metastable phases in many of these materials (e.g., the hidden state in $1T$-$TaS_2$(*4, 6, 10*)) has been a topic of great interest. However, methods to prepare, detect and study metastable states are still under development. Indeed, whether such states are present in a wide range of materials is still an open question.

Ultrafast spectroscopies are emerging as powerful tools to study correlated materials because of their unique abilities to probe the dominant microscopic interactions. In particular, time- and angle-resolved photoemission spectroscopy (trARPES), time-resolved magneto-optical Kerr spectroscopy and time-resolved electron/x-ray diffraction can probe the coupled dynamics of the charge, spin and lattice systems, respectively, after a material is excited by an ultrafast laser pulse(*13*). For example, many past studies explored how the energy flows from the excited electron and spin systems to the phonon bath and how the state of material evolves(*14–19*). Other work uncovered how the characteristic timescales for the insulator to metal phase transition depend on different CDW interactions(*20*). More recently, a combination of trARPES and x-ray diffraction was used to directly measure the deformation potential(*21*). Using tabletop high harmonic generation, we used trARPES to identify a new super-excited magnetic state where the spin system absorbs sufficient energy within 20 fs to subsequently proceed through a magnetic phase transition(*22*).

Here we present a new ultrafast electron calorimetry technique that can systematically uncover new phases of matter. Using trARPES, we measure the electron temperature, band structure and heat capacity. We then show that this information can be used as a very sensitive probe of where phase changes occur within quantum materials, because electrons react very quickly — on femtosecond to attosecond timescales. Moreover, their heat capacity is generally the smallest component of the total heat capacity (×100 less than the lattice), and cannot be measured under thermal equilibrium conditions. We use our new ultrafast electron calorimetry technique to uncover a new charge-ordered phase in the transition metal dichalcogenide 1$T$-TaSe$_2$ that is simple to excite and that has remarkable properties. As the laser excitation fluence is increased to completely melt the CDW order, we observe a significant reduction in the effective heat capacity, by 70%. This indicates that the electron-phonon coupling abruptly switches from nearly homogeneous to mode-selective. The strong inhomogeneity within the phonon system, i.e., where some strongly coupled phonon modes thermalize to above the transition temperature ($T_c$) while the rest do not, drives the material into a new long-lived metastable CDW state, as illustrated in Figure 1 and Movie S1. This metastable CDW phase, with a continuously tunable order parameter depending on the laser fluence, is stable over timescales ranging from picoseconds to hundreds of picoseconds, thus dramatically enriching the landscape of metastable ordered phases of 1$T$-TaSe$_2$. Most significantly, this ultrafast phase transformation requires significantly less energy (~ 60%) than is required to melt the CDW order under thermal equilibrium conditions. Moreover, it has a very simple excitation mechanism that can still result in very selective excitation of a subset of phonon modes. Our findings thus uncover new manifestations of electron-phonon coupling and phase transition pathways in quasi-two-dimensional CDW systems. Finally, we note that our approach is general, and can be used to uncover the presence of hidden phases in other materials.

**Results**

1$T$-TaSe$_2$ crystallizes in a quasi-two-dimensional crystal structure consisting of an undistorted hexagonal Ta layer sandwiched between two Se layers in the high-temperature phase. At lower temperatures, 1$T$-TaSe$_2$ transforms to an incommensurate CDW phase at 600 K and then to a commensurate CDW phase (first order transition) at $T_c$ = 470 K(*23*). This CDW phase is characterized by an electron density modulation and corresponding lattice distortion in a $\sqrt{13} \times \sqrt{13}$ periodicity(*23*). Each supercell forms a 13-atom star-of-David cluster, where 12 outer atoms are displaced inward, as shown in the left panel of Figure 1. In our trARPES experiment, the sample in the CDW state (300 K) is photo-excited by a 1.6 eV pump pulse, then the electronic structure as a function of time delay and pump fluence is measured by a 22.4 eV probe pulse produced from high harmonic generation(*24*).

The band structure of 1$T$-TaSe$_2$ in the CDW state has prominent features including band folding and an opening of an energy gap, that can be clearly seen in the photoemission spectrum along the Γ-M direction before laser excitation. The disappearance of these features after laser excitation suggests that the CDW order is destroyed by an ultrafast laser pulse, as shown in Fig. 2A and also studied extensively in past work(*14, 16, 25–27*). During this process, the electron system is directly excited to a non-thermal energy distribution, which rapidly thermalizes to a Fermi-Dirac distribution. To better understand the ultrafast CDW phase transition, we extract the time evolution of the electron temperature $T_e$ by fitting energy distribution curves of the ARPES spectra with a Fermi-Dirac function multiplied by a Lorentzian-shape density of states (see Supporting Online Materials (SOM)). The results are displayed in Figs. 2B and S4, which show that the electrons reach a maximum temperature of thousands of Kelvin. To investigate the effect of the spatial charge redistribution at high temperature, we perform density functional theory (DFT) calculations on 1$T$-TaSe$_2$. As shown in Fig. 2C and Fig. S3, at room temperature the charge is mainly distributed near the center of the star-of-David in real-space (around the inner Ta atoms). In contrast, after laser excitation, the elevated electron temperature smears-out the charge localization in the center of each star due to significant excitation of electrons to above the Fermi level ($E_F$). This subsequently modifies the interatomic potential and initiates rapid lattice rearrangement. On the short timescales required for this structural rearrangement to be launched (20–100 fs), the lattice temperature remains near 300 K.

In order to understand the complex interactions in the material, it is useful to investigate energy flow and coupling between the electron and lattice degrees of freedom. We first analyze the relaxation of the electron temperature as a function of laser fluence, which is governed by the electron-phonon coupling(*13*). As shown in Fig. 2D, the electron temperature decays to a plateau by ~ 4 ps, indicating that the electron bath has equilibrated with at least a part of the phonon bath. This quasi-equilibrium temperature increases with the laser fluence, and surprisingly, shows an abrupt increase at a critical fluence ($F_c$) of about 0.7 mJ/cm$^2$, corresponding to when the electron temperature and part of the phonon bath cross the transition temperature $T_c$ (see Fig. 2E). Note that in many material systems, the electrons and lattice (i.e., all the phonons) would have equilibrated by 4 ps. However, the solid red curve in Fig. 2E plots the quasi-equilibrium temperature calculated from the equilibrium heat capacity of the material (see SOM), showing that this is only true for $F < F_c$ in 1$T$-TaSe$_2$. The jump in temperature at $F_c$ (Fig. 2E) thus reveals that there is a sudden reduction in the effective lattice heat capacity for $F > F_c$. This reduction is as large as 70% compared to below $F_c$, as further confirmed by the increased slope of the fluence-dependent electron temperature (see the red dashed line in Fig. 2E). In the temperature range explored here, the heat capacity is close to the Dulong-Petit limit and thus cannot change so dramatically under equilibrium conditions. Therefore, our results suggest that only a subset of phonon modes, that are strongly coupled to the hot electrons, contribute to the effective heat capacity at $F > F_c$, as illustrated in Fig. 1.

Given the anomalous changes in electron-phonon coupling observed at $F_c$, we also investigate the dynamic band structure. Here we focus on the Ta $5d$ band near $E_F$, as shown in Fig. 2A. The time-dependent photoemission spectrum in Fig. 3A clearly shows the oscillation of the band position (in binding energy), that is coupled to the excited CDW amplitude mode (breathing mode of the stars)(*14*, *25*, *28*). Figure 3B plots the instantaneous band shift after excitation for two typical fluences. For low laser fluences in the perturbative regime, the band shift and electron temperature are strongly correlated. However, once the laser fluence exceeds $F_c$, the CDW order melts within a half cycle of a strongly damped amplitude mode oscillation, before entering the metastable phase. We can model this behavior as a classical harmonic oscillator(*29*), with an equation of motion for the band position $Q(\Delta t)$ given by

$$\ddot{Q}(\Delta t) = -\omega_0^2[Q(\Delta t) - Q_0(\Delta t)] - 2\gamma \dot{Q}(\Delta t) \tag{1}$$

where $\omega_0$ is the angular frequency, and $\gamma$ is the damping constant. The quasi-equilibrium coordinate $Q_0(\Delta t)$ is taken as

$$Q_0(\Delta t) = C_1(1 - e^{-\Delta t/\tau_1}) + C_2(e^{-\Delta t/\tau_2} - 1) \tag{2}$$

where $\tau_1$ and $\tau_2$ are the exponential rise and decay time constants. The data fit well to this model (Fig. 3B) and we plot in Fig. 3C the extracted oscillation frequencies and damping constants for different laser fluences. The oscillation frequency is ~ 2 THz and decreases slightly with fluence, while the damping constant increases linearly and interestingly, exceeds the oscillation frequency near $F_c$.

Since the band shift is strongly coupled to both the atomic displacement (structural order parameter) and the band folding or energy gap (electronic order parameter, see SOM), it can represent the CDW order in this material. Next, we study the time evolution of the band shift as a function of excitation fluence to gain more insight into the ultrafast CDW transition.

Figure 3D plots the peak of the quasi-equilibrium coordinate $Q_0$, which represents the ultrafast CDW order suppression (orange curve), as well as the band shift at a delay of 4 ps (blue curve). At the same critical fluence $F_c$ observed in the electron temperature dynamics (Fig. 2E), we observe that the peak of $Q_0$ saturates at the expected thermal equilibrium value of ~ 175 meV — corresponding to complete closure of the gap, while the band shift at 4 ps shows a kink (turning point). It is worth mentioning that the saturation of $Q_0$ is also consistently reflected in the fluence dependent maximum $T_e$, which suggests a change of the effective electronic heat capacity at $F_c$ (see SOM). These results reveal a close relationship between the electron-phonon coupling and the CDW transition, and suggest that once the CDW order is completely melted at $F > F_c$, the phase of the material changes in an extraordinary way. Indeed, the band shift after reaching a quasi-equilibrium at 4 ps behaves as if it undergoes a continuous phase transition (Fig. 3D), which is in strong contrast to the well-identified first order transition that occurs under thermal equilibrium conditions. Moreover, this new state lasts for hundreds of picoseconds and thus provides access to a metastable CDW order (see SOM) in $1T$-TaSe$_2$. As shown in Fig. 4, that plots the band shift (order parameter) as a function of both time delay and laser fluence, this new state does not correspond to either of the equilibrium phases of the material. Note that this band shift dramatically highlights the sudden change in the order parameter and material phase that occurs at the critical fluence.

Under weak excitation ($F < F_c$), our data show strong initial electron-phonon coupling (Fig. 3E) followed by nearly homogeneous excitation of the entire phonon bath within 4 ps (Fig. 2E). Moreover, the original CDW order recovers on comparable timescales to the decay of the electron temperature (Fig. 3E). In contrast, as the laser excitation fluence exceeds $F_c$, the averaged electron-phonon coupling constant increases within the first picoseconds (Fig. 3F and G), likely related to the enhancement of the density of states at $E_F$ as the flat band near the Brillouin zone center shifts upwards. This is accompanied by a sudden reduction in the effective lattice heat capacity at $F_c$ — the electron-phonon coupling switches to being mode-selective, and only a subset of phonons are coupled to the hot electrons and thermalized above $T_c$. This strong inhomogeneity within the phonon system has no counterpart under thermal equilibrium conditions and pushes the material into a new metastable CDW state (see Movie S1 and Fig. S8) along a path that lowers the free energy. This process takes longer for stronger laser excitation, and no longer tracks the decay of the electron temperature (Fig. 3F and G).

Thus, the origin of the new long-lived metastable state can be clearly attributed to the strong mode-selectivity of the electron-phonon coupling, as illustrated in Figs. 1 and 4. We note that anisotropic electron-phonon coupling has also been reported in cuprates(*30*, *31*), graphene(*32*) and theoretical calculation on metals(*33*). By investigating electron-phonon coupling in both the time domain and as a function of laser fluence, the results discussed here provide even more direct evidence of anisotropy, and in addition, reveal the transition from nearly homogeneous to mode-selective electron-phonon coupling and relate it to the formation of a new long-lived metastable state. Taking advantage of this kind of selective excitation of phonons, our data suggest that ultrashort laser pulses are able to transform the material into the new and otherwise unreachable intermediate CDW state. And quite importantly, the laser energy required for the phase transition through this non-equilibrium pathway is significantly lower (~ 60%, see SOM) than that under thermal equilibrium conditions. Hence, this experiment demonstrates an ultrafast, controllable and energetically efficient way to manipulate the state of 1$T$-TaSe$_2$, which can likely be extended to many other materials.

**Discussion**

Finally, we discuss the underlying physical mechanisms that may be giving rise to our observations. In thermal equilibrium for a CDW material such as 1$T$-TaSe$_2$, it has been pointed out that the phonon frequency softens around the CDW wave vector $\mathbf{q}_{CDW}$ (known as a Kohn anomaly)(*34*). For temperatures below $T_c$, the frequency of some modes around $\mathbf{q}_{CDW}$ approaches zero (giving rise to an imaginary value in *ab initio* calculations signaling the instability of the spatial configuration) and this induces a lattice instability. In the case of ultrafast excitation of 1$T$-TaSe$_2$, once the CDW order is suppressed, the zero phonon frequency recovers to a real value close to zero at the phase transition (i.e., at $F = F_c$). The electron-phonon coupling constant for a specific phonon mode (phonon branch $v$ at wave vector **q**) is

$$\lambda_{\mathbf{q}v} = \frac{\gamma_{\mathbf{q}v}}{\pi N(E_F)\omega_{\mathbf{q}v}^2} \qquad (3)$$

where $\gamma_{\mathbf{q}v}$ is the phonon linewidth, $N(E_F)$ is the density of states per spin and $\omega_{\mathbf{q}v}$ is the phonon frequency. The ultralow frequency and large linewidth(*35–37*) of some phonon modes can lead to a mode-selective enhancement of the electron-phonon coupling to breathing and associated modes. This would explain the enhanced cooling of the electron bath on early 1 ps timescales, as well as the reduction in the heat capacity of the lattice — since some other phonon modes are more weakly coupled to the electron system. This might also be explained by the material becoming more 2D-like in the metastable state or due to surface decoupling. However, we see no evidence of surface decoupling in the ARPES spectra. A more quantitative explanation of the exotic electron-phonon coupling behavior during this ultrafast phase transition will require more advanced microscopic theories.

In summary, we use ultrafast electron calorimetry to measure the dynamic electron temperature, band structure and heat capacity as a function of laser fluence, and demonstrate that it is a powerful method for not only characterizing important physical quantities (e.g., electron-phonon coupling) and transient excited states, but also for revealing when and where microscopic interactions change the state of a material. We then use it to uncover a new long-lived metastable state that is mediated by mode-selective electron-phonon coupling. This new state enriches the phase diagram of 1$T$-TaSe$_2$, and thus allows for better understanding and manipulation of CDW and electron-phonon interactions. Thus, time- and angle-resolved photoemission can uncover a broad set of exotic phenomena in correlated materials, making it possible to uncover novel phases that are elusive to traditional spectroscopies.

## Materials and Methods

**Experiments.** The trARPES experiments were done in a pump-probe scheme. The 1.6 eV infrared (IR) laser beam with a pulse duration of ~ 30 fs was generated by a Ti:Sapphire oscillator-amplifier system (KMLabs Dragon) at a repetition rate of 4 kHz. It was then split into pump and probe lines. In the probe line, the beam was first frequency doubled by a β-Barium borate crystal to 3.2 eV and then focused into a waveguide filled with Kr gas for high harmonic generation. The data reported here were recorded using 22.4 eV photons (7$^{th}$ order). A delay stage in the pump line was used to control the time delay between the pump and probe pulses. Single crystals of 1$T$-TaSe$_2$ were cleaved *in-situ* and measured at 300 K, under a vacuum of $3 \times 10^{-10}$ Torr. The photoelectrons were detected by a SPECS PHOIBOS 100 energy analyzer. The overall energy resolution is about 130 meV, which is mainly limited by the band width of the ultrashort laser pulses.

**Theoretical calculations.** Our DFT calculations employ the projector augmented wave (PAW) method encoded in Vienna *ab initio* simulation package (VASP). The PAW method is used to describe the wavefunctions near the core, and the generalized gradient approximation within the Perdew-Burke-Ernzerhof (PBE) parameterization is employed as the electron exchange-correlation functional.

**References and Notes**

1. D. Fausti *et al.*, Light-Induced Superconductivity in a Stripe-Ordered Cuprate. *Science*. **331**, 189–191 (2011).
2. Y. H. Wang, H. Steinberg, P. Jarillo-Herrero, N. Gedik, Observation of Floquet-Bloch States on the Surface of a Topological Insulator. *Science*. **342**, 453–457 (2013).
3. R. Mankowsky *et al.*, Nonlinear lattice dynamics as a basis for enhanced superconductivity in YBa2Cu3O6.5. *Nature*. **516**, 71–73 (2014).
4. L. Stojchevska *et al.*, Ultrafast Switching to a Stable Hidden Quantum State in an Electronic Crystal. *Science*. **344**, 177–180 (2014).
5. V. R. Morrison *et al.*, A photoinduced metal-like phase of monoclinic VO2 revealed by ultrafast electron diffraction. *Science*. **346**, 445–448 (2014).
6. I. Vaskivskyi *et al.*, Controlling the metal-to-insulator relaxation of the metastable hidden quantum state in 1T-TaS2. *Sci. Adv.* **1**, e1500168 (2015).
7. M. Mitrano *et al.*, Possible light-induced superconductivity in K3 C60 at high temperature. *Nature*. **530**, 461–464 (2016).
8. J. Zhang *et al.*, Cooperative photoinduced metastable phase control in strained manganite films. *Nat. Mater.* **15**, 956–960 (2016).
9. D. N. Basov, R. D. Averitt, D. Hsieh, Towards properties on demand in quantum materials. *Nat. Mater.* **16**, 1077–1088 (2017).
10. K. Sun *et al.*, Hidden CDW states and insulator-to-metal transition after a pulsed femtosecond laser excitation in layered chalcogenide 1T-TaS2−xSex. *Sci. Adv.* **4**, eaas9660 (2018).
11. G. Grüner, The dynamics of charge-density waves. *Rev. Mod. Phys.* **60**, 1129–1181 (1988).
12. D. J. Scalapino, A common thread: The pairing interaction for unconventional superconductors. *Rev. Mod. Phys.* **84**, 1383–1417 (2012).
13. C. Giannetti *et al.*, Ultrafast optical spectroscopy of strongly correlated materials and high-temperature superconductors: a non-equilibrium approach. *Adv. Phys.* **65**, 58–238 (2016).
14. L. Perfetti *et al.*, Time Evolution of the Electronic Structure of 1T-TaS2 through the Insulator-Metal Transition. *Phys. Rev. Lett.* **97**, 67402 (2006).
15. F. Schmitt *et al.*, Transient Electronic Structure and Melting of a Charge Density Wave in TbTe3. *Science*. **321**, 1649–1652 (2008).
16. M. Eichberger *et al.*, Snapshots of cooperative atomic motions in the optical suppression of charge density waves. *Nature*. **468**, 799–802 (2010).

17. T. Rohwer *et al.*, Collapse of long-range charge order tracked by time-resolved photoemission at high momenta. *Nature*. **471**, 490–494 (2011).
18. C. L. Smallwood *et al.*, Tracking Cooper Pairs in a Cuprate Superconductor by Ultrafast Angle-Resolved Photoemission. *Science*. **336**, 1137–1139 (2012).
19. F. Boschini *et al.*, Collapse of superconductivity in cuprates via ultrafast quenching of phase coherence. *Nat. Mater.* **17**, 416–420 (2018).
20. S. Hellmann *et al.*, Time-domain classification of charge-density-wave insulators. *Nat. Commun.* **3**, 1069 (2012).
21. S. Gerber *et al.*, Femtosecond electron-phonon lock-in by photoemission and x-ray free-electron laser. *Science*. **357**, 71–75 (2017).
22. P. Tengdin *et al.*, Critical behavior within 20 fs drives the out-of-equilibrium laser-induced magnetic phase transition in nickel. *Sci. Adv.* **4**, eaap9744 (2018).
23. J. A. Wilson, F. J. Di Salvo, S. Mahajan, Charge-Density Waves in Metallic, Layered, Transition-Metal Dichalcogenides. *Phys. Rev. Lett.* **32**, 882–885 (1974).
24. A. Rundquist *et al.*, Phase-Matched Generation of Coherent Soft X-rays. *Science*. **280**, 1412–1415 (1998).
25. J. C. Petersen *et al.*, Clocking the Melting Transition of Charge and Lattice Order in 1T-TaS2 with Ultrafast Extreme-Ultraviolet Angle-Resolved Photoemission Spectroscopy. *Phys. Rev. Lett.* **107**, 177402 (2011).
26. C. Sohrt, A. Stange, M. Bauer, K. Rossnagel, How fast can a Peierls–Mott insulator be melted? *Faraday Discuss.* **171**, 243–257 (2014).
27. S. Sun *et al.*, Direct observation of an optically induced charge density wave transition in 1T-TaSe2. *Phys. Rev. B*. **92**, 224303 (2015).
28. J. Demsar, L. Forró, H. Berger, D. Mihailovic, Femtosecond snapshots of gap-forming charge-density-wave correlations in quasi-two-dimensional dichalcogenides 1T-TaS2 and 2H-TaSe2. *Phys. Rev. B*. **66**, 41101 (2002).
29. T. K. Cheng *et al.*, Mechanism for displacive excitation of coherent phonons in Sb, Bi, Te, and Ti2O3. *Appl. Phys. Lett.* **59**, 1923–1925 (1991).
30. L. Perfetti *et al.*, Ultrafast Electron Relaxation in Superconducting Bi2Sr2CaCu2O8+δ by Time-Resolved Photoelectron Spectroscopy. *Phys. Rev. Lett.* **99**, 197001 (2007).
31. S. Dal Conte *et al.*, Disentangling the Electronic and Phononic Glue in a High-Tc Superconductor. *Science*. **335**, 1600–1603 (2012).
32. J. C. Johannsen *et al.*, Direct View of Hot Carrier Dynamics in Graphene. *Phys. Rev. Lett.* **111**, 27403 (2013).
33. P. Maldonado, K. Carva, M. Flammer, P. M. Oppeneer, Theory of out-of-equilibrium

ultrafast relaxation dynamics in metals. *Phys. Rev. B*. **96**, 174439 (2017).

34. W. Kohn, Image of the Fermi Surface in the Vibration Spectrum of a Metal. *Phys. Rev. Lett.* **2**, 393–394 (1959).

35. F. Weber *et al.*, Extended Phonon Collapse and the Origin of the Charge-Density Wave in 2H-NbSe2. *Phys. Rev. Lett.* **107**, 107403 (2011).

36. X. Zhu, Y. Cao, J. Zhang, E. W. Plummer, J. Guo, Classification of charge density waves based on their nature. *Proc. Natl. Acad. Sci.* **112**, 2367–2371 (2015).

37. Y. Liu *et al.*, Nature of charge density waves and superconductivity in 1T-TaSe2-xTex. *Phys. Rev. B*. **94**, 45131 (2016).

**Acknowledgements:** We thank James K. Freericks for useful discussions. **Funding:** M.M. and H.K. gratefully acknowledge support for the ARPES measurements from the National Science Foundation through the JILA Physics Frontiers Center PHY-1125844, and a Gordon and Betty Moore Foundation EPiQS Award GBMF4538. **Author contributions:** X.S., W.Y., Y.Z. and Z.T. performed trARPES measurements. K.R. provided 1$T$-TaSe$_2$ samples. X.S., W.Y. and Y.Z. analyzed the data. X.W., R.T. and P.M.O. provided theoretical input. X.S., K.R., M.B., H.K. and M.M. wrote the manuscript with inputs from all other authors. X.S., W.Y., Y.Z. H.K. and M.M. conceived the investigation. All authors discussed the underlying physics. **Competing interests:** H.K. and M.M. have a financial interest in a laser company, KMLabs, that produces the lasers and HHG sources used in this work. H.K. is partially employed by KMLabs. The authors declare that they have no other competing interests. **Data and materials availability:** All data needed to evaluate the conclusions in the paper are present in the paper and/or the Supplementary Materials. Additional data related to this paper may be requested from the authors.

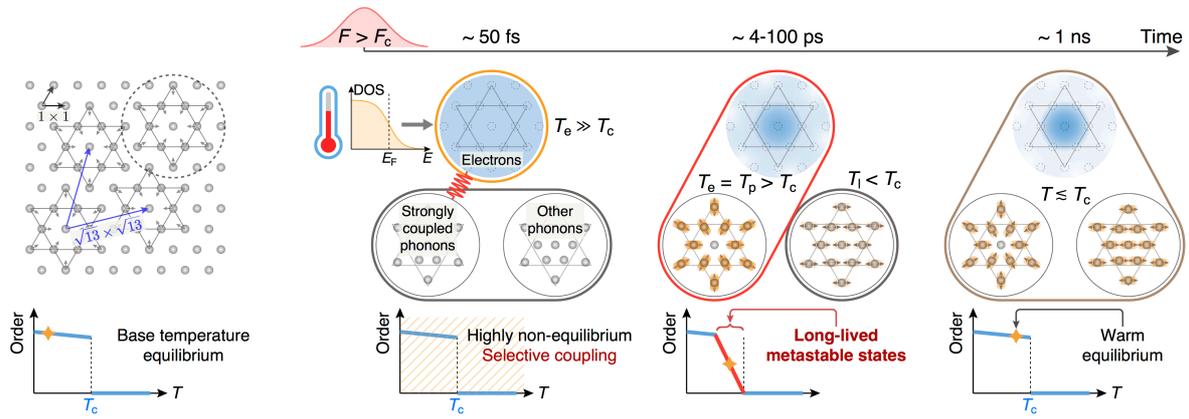

**Fig. 1. Ultrafast electron calorimetry can measure the dynamic electron temperature and band structure, to uncover a new long-lived metastable state mediated by mode-selective electron-phonon coupling.** The upper left panel shows the top view of the Ta plane in 1$T$-TaSe$_2$. In the CDW state, displacement of the Ta atoms leads to a $\sqrt{13} \times \sqrt{13}$ superstructure consisting of 13-atom star-of-David clusters. After laser excitation, the evolution of the sample is determined first by the electron temperature and then by electron-phonon coupling, which depend on the fluence. For strong laser excitation, the electron-phonon coupling switches from nearly homogeneous to mode-selective. The resulting inhomogeneity within the phonon bath drives the material into a new long-lived metastable CDW state. The blue shading represents the electron density in the real-space, the grey circles represent Ta atoms, both amplitudes are exaggerated for better visualization. $T_e$, $T_p$ and $T_l$ refer to the temperatures of the electron, strongly-coupled phonons, and the rest of the phonon bath, respectively.

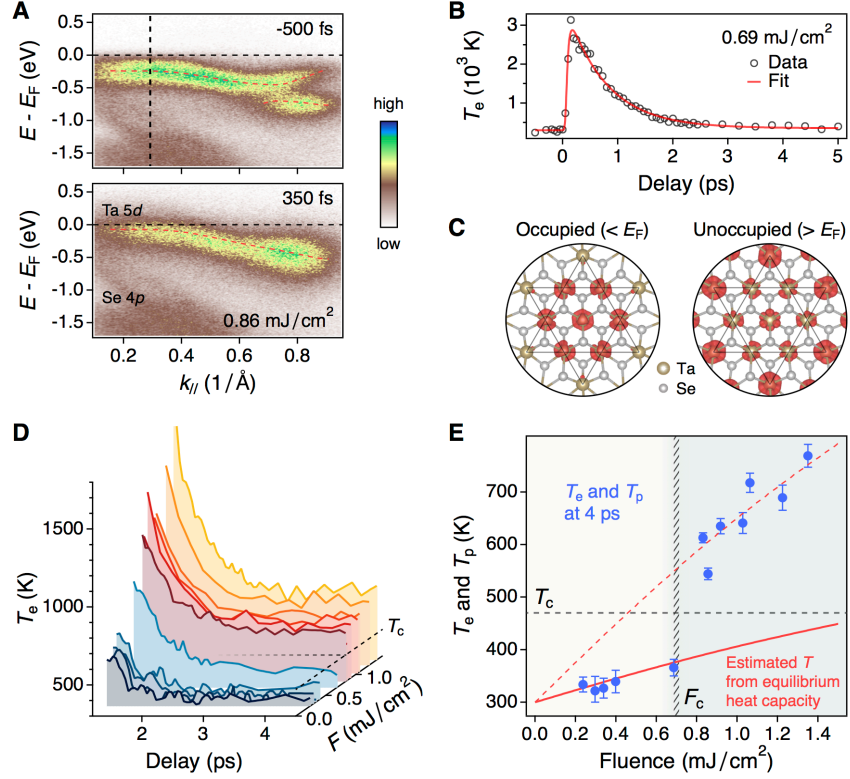

**Fig. 2. Evolution of the electron temperature and sudden change in the electron-phonon coupling.** (**A**) Photoemission spectra along the Γ-M direction before and after (350 fs) laser excitation with a fluence of 0.86 mJ/cm². The disappearance of band folding and the CDW gap clearly suggest the collapse of the CDW order. (**B**) Temporal evolution of the electron temperature $T_e$ at a fluence of 0.69 mJ/cm². The red curve is the two-exponential fit to the data. (**C**) Calculated charge densities (red) in a star-of-David integrated over the energy windows on the occupied ($E_F - 1$ eV, $E_F$) and unoccupied ($E_F + 0.2$ eV, $E_F + 1.2$ eV) sides. (**D**) Electron temperature dynamics as a function of laser fluence. (**E**) $T_e$ at 4 ps, when the electron bath is nearly in equilibrium with part of the phonon bath ($T_p$). The error bars represent the measurement uncertainties. For $F < F_c$, this quasi-equilibrium temperature reaches the expected value in thermal equilibrium, as indicated by the solid red curve. However, for $F > F_c$, the temperature abruptly increases above the red curve, indicating a step decrease in the effective heat capacity — only a subset of phonons are strongly coupled to the hot electrons.

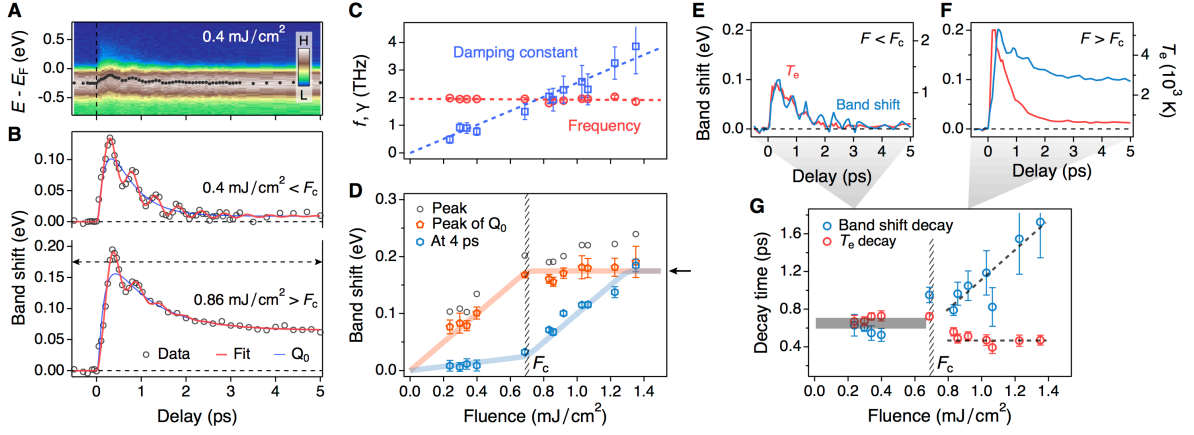

**Fig. 3. Electronic band dynamics and the metastable state.** (**A**) Time-dependent photoemission spectrum at the momentum $k_{//}$ as indicated by the vertical dashed line in Fig. 2A. The black dots represent the band positions. (**B**) Energy shift of the band shown in (A) at two representative laser fluences. Red curves indicate fits of the data, blue curves indicate the quasi-equilibrium coordinate $Q_0$ as described in main text. (**C**) Extracted oscillation frequency $f$ and damping constant $\gamma$ as a function of fluence. (**D**) Peak of band shift, peak of $Q_0$ and band shift at 4 ps as a function of fluence. The arrow indicates the saturation value corresponding to melting of the CDW order. At fluences higher than 0.7 mJ/cm$^2$, the material evolves into a new long-lived metastable state. (**E** and **F**) Comparison between the evolution of $T_e$ and the band shift, as a function of fluence. (**G**) Decay timescales of $T_e$ and the band shift as a function of fluence. When $F > F_c$, the faster decay of $T_e$ indicates an enhancement in the averaged electron-phonon coupling, while the decay of band shift starts to deviate from the former and becomes slower. The error bars include the measurement uncertainties and the standard deviation of the fitting.

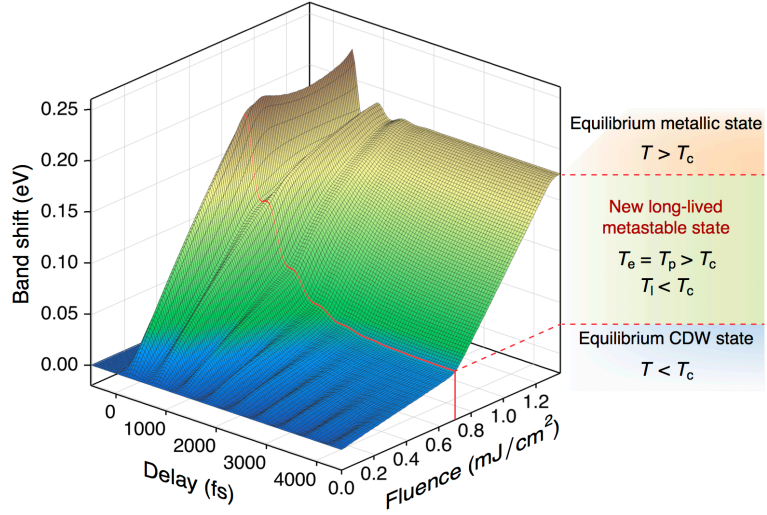

**Fig. 4. Experimental band shift (order parameter) as a function of time delay and laser fluence.** At fluences higher than $F_c$ (~ 0.7 mJ/cm$^2$ indicated by the red line), the material suddenly evolves into a new metastable state that is distinct from either of the equilibrium phases. $T_e$, $T_p$ and $T_l$ refer to the temperatures of the electron, strongly-coupled phonons, and the rest of the phonon bath, respectively.